\documentstyle[prl,aps]{revtex}
\begin{document}
\draft
\tighten
\onecolumn
%\twocolumn[\hsize\textwidth\columnwidth\hsize\csname @twocolumnfalse\endcsname
\title{Phonon Squeezed States: Quantum Noise Reduction in Solids}
\author{Xuedong Hu$^1$ and Franco Nori$^2$}
\address{
1 Department of Physics, University of Maryland, College Park, MD 20742
\\
2 Department of Physics, The University of Michigan, Ann Arbor,
Michigan 48109-1120
}
\date{\today}
\maketitle
\begin{abstract}
This article discusses quantum fluctuation properties of a
crystal lattice, and in particular, phonon squeezed states.
Squeezed states of phonons allow a reduction in
the quantum fluctuations of the atomic displacements
to below the zero-point quantum noise level of
coherent phonon states.
Here we discuss our studies
of both continuous-wave and impulsive second-order Raman
scattering mechanisms.  The later approach was used to
experimentally suppress (by one part in a million)
fluctuations in phonons.
We calculate the expectation values and fluctuations of both the
atomic displacement and the lattice amplitude operators, as well as
the effects of the phonon squeezed states on macroscopically
measurable quantities, such as changes in the dielectric constant.
These results are compared with recent experiments.
Further information, including preprints and animations, are available
in http://www-personal.engin.umich.edu/\~{}nori/squeezed.html
\end{abstract}
%\vspace*{-0.1in}
%\pacs{PACS numbers: 42.50.Dv, 42.50.Lc, 42.50.Ct, 42.65.Dr}
%
%\vskip2pc]
%\narrowtext
\vspace*{+0.2in}
\section{Introduction}
%{\it Introduction}.---
Classical phonon optics \cite{phononopt} has succeeded in producing
many acoustic analogs of {\it classical optics}, such as phonon
mirrors, phonon lenses, phonon filters, and even ``phonon microscopes''
that can generate acoustic pictures with a resolution comparable to
that of visible light microscopy.  Most phonon optics experiments use
heat pulses or superconducting transducers to generate {\em incoherent}
phonons, which propagate ballistically in the crystal.  These ballistic
incoherent phonons can then be manipulated by the above-mentioned
devices, just like in geometric optics.
Phonons can also be excited {\em phase-coherently}.  For instance,
coherent acoustic waves with frequencies of up to $10^{10}$ Hz can be
generated by piezoelectric oscillators.  Lasers have also been used to
generate coherent acoustic and optical phonons through stimulated
Brillouin and Raman scattering experiments.  Furthermore, in recent
years, it has been possible to track the phases of coherent optical
phonons \cite{review}
due to the availability of femtosecond-pulse
ultrafast lasers (with a pulse duration shorter than a phonon period),
%\cite{ultra},
and techniques that can measure optical reflectivity with
accuracy of one part in $10^{6}$.
In most situations involving phonons, a {\it classical} description is
adequate.  However, at low enough temperatures, {\it quantum}
fluctuations become dominant.  For example, a recent study \cite{fnori}
shows that quantum fluctuations in the atomic positions can indeed
influence observable quantities (e.g., the Raman line shape) even when
temperatures are not very low.  With these facts in mind, and prompted
by the many exciting developments in {\it classical} phonon optics,
coherent phonon experiments, and (on the other hand) squeezed states of
light \cite{special}, we would like to explore phonon analogs of {\it
quantum} optics. In particular, we study the dynamical and quantum
fluctuation properties of the atomic displacements, in analogy with the
modulation of quantum noise in light.  Specifically, we study
single-mode and two-mode phonon coherent and squeezed states, and then
focus on a polariton-based approach to achieve smaller quantum noise
than the zero-point fluctuations of the atomic lattice.
The concepts of coherent and squeezed states were both originally
proposed in the context of quantum optics.  A coherent state is a
phase-coherent sum of number states.  In this state, the quantum
fluctuations in any pair of conjugate variables are at the lower limit
of the Heisenberg uncertainty principle.  In other words, a coherent
state is as ``quiet'' as the vacuum state.  Squeezed states
\cite{special} are interesting because they can have {\it smaller
quantum noise than the vacuum state} in one of the conjugate variables,
thus having a promising future in different applications ranging from
gravitational wave detection to optical communications.  In addition,
squeezed states form an exciting group of states and can provide unique
insight into quantum mechanical fluctuations.
In recent years, squeezed states are also being explored in a variety
of non-quantum-optics systems, including ion-motion and classical
squeezing \cite{Rugar}, molecular vibrations \cite{Janszky}, polaritons
\cite{Birman,xhuprb},
%phonons in superfluid $^3$He \cite{McKenzie},
and phonons in crystals
\cite{xhuaps,xhurmp,xhuprl,xhuprl97,AIPnews}.
Ref.~\cite{xhurmp,xhuprl,xhuprl97} propose a second-order Raman
scattering (SORS) process for phonon squeezing:  if the two incident
light beams are in coherent states, the phonons generated by the SORS
are in a two-mode squeezed state.
Here we first present an introduction to the subject of coherent and
squeezed phonons, and later on consider both the continuous wave case
studied in \cite{xhurmp,xhuprl97} and the impulsive case studied in
\cite{xhuprl97,Garrett}.
Squeezed phonons could be detected by measuring the intensity of the
reflected or transmitted probe light
\cite{xhurmp,xhuprl,xhuprl97,Garrett}.  This method has been used to
detect coherent phonon amplitudes, since reflectivity and transmission
are closely related to the atomic displacements in a crystal.
Measuring a transmitted probe light pulse, Ref.~\cite{Garrett} observed
squeezed phonons produced by an impulsive SORS. The intensity of the CW
SORS signal for many materials might be too weak to be detected with
current techniques, but might be accessible in the future.
\section{Analogies and Differences between Phonons and Photons}
\label{sct:analogy}
Coherent and squeezed states were initially introduced to describe
photons.  Here we are interested in applying these concepts to
phonons.  Although both photons and phonons are bosons, they do have
significant differences, and the physics of squeezed states of light
cannot be straightforwardly extended to phonons.
Table I is a chart presenting a schematic comparison between phonons
and photons.  Below we briefly mention a few important similarities and
differences that are relevant to our study.

Photons are elementary particles with no internal structure, thus are
sometimes called simple bosons.  On the other hand, phonons describe
the collective displacements of very many atoms in a crystal, and are
thus sometimes described as composite bosons \cite{Kohn}.  Phonons are
bosons because of the commutation relation between the coordinate and
momentum operators.  Kohn and Sherrington \cite{Kohn} pioneered the
research on composite bosons like phonons, excitons, etc., and
classified them into two categories, with type-I referring to those
bosons composed of an even number of fermions (such as $^4$He atoms),
and type-II referring to those that are collective excitations---such
as phonons, excitons, magnons, etc.  In this sense it is also possible
to consider photons to be type-II composite bosons \cite{Kohn}, because
they are the energy quanta of electromagnetic field modes.  Their
commutation relation originates from the simple harmonic oscillators
that are used to quantize the electromagnetic field.  Essentially, both
photons and phonons are field quanta: photons are quanta of a
continuous field, while phonons are quanta of a discrete field.
Non-interacting phonons are used to describe harmonic crystal
potentials.  However, anharmonicity, which leads to phonon-phonon
interactions, is always present.  Some properties of solids, such as
lattice heat conductivity and thermal expansion, solely depend on the
anharmonic terms in the crystal potential.  In other words, phonons in
general interact with each other.  For photons, the situation is
somewhat different.  In vacuum and at low intensity, photon
interactions are so weak that the rule of linear superposition holds.
However, in nonlinear media, photons are effectively interactive, with
their interaction mediated by the atoms.
As mentioned above, phonons exist in discrete media.  Therefore,
phonons have cut-off frequencies, which put an upper-limit to their
energy spectra.  For a diatomic lattice, this limit is of the order of
0.1 eV, which is in the infrared region.  Photons, on the other hand,
do not have such an upper bound for their energy.  In addition, the
discrete atomic lattice and the massive atoms lead to a finite
zero-point fluctuation in the phonon field, while the continuous photon
modes and the massless photons contribute to a divergent zero-point
fluctuation in the photon field.
The dispersion relations for photons and phonons are qualitatively
different.  Photons in free space have a linear dispersion relation.
On the other hand, phonons have complicated nonlinear dispersion
relations which generally have several acoustic and optical branches.
The acoustic branches are linear around the center of the first
Brillouin zone, i.e., the ${\bf k} = 0$ point, which is at the
continuum limit.  When the quasi-wave-vector ${\bf k}$ is close to the
first Brillouin zone boundary, $\omega$ saturates.  The optical
branches of the phonons have a different profile.  Their dispersion
relations are flat near ${\bf k} = 0$, where $\omega = \omega_0$.
Furthermore, as ${\bf k}$ increases, $\omega$ decreases; indeed, the
optical phonon dispersion relation can be even more complicated
depending on the lattice structure.  Compared to photons, which
generally have relatively simple dispersion relations, phonons have
nonlinear dispersion relations that make it more difficult to satisfy
both energy and momentum (in fact, quasi-momentum) conservation laws
simultaneously.
The order of magnitude of the crystal cohesion energy determines
that phonons have very low energies.  In addition, phonons can easily
couple to many other excitations which are in a similar energy range,
and be perturbed by thermal fluctuations even at low temperatures.  All
these couplings make phonon dynamics very dissipative.  Due to its
strong damping, coherent phonons have very short lifetimes ($\sim 50$
picosecond for optical phonons, while larger for acoustic phonons)
\cite{review}.  On the other hand, there exist many materials in which
photons can propagate with little dissipation, and furthermore very
long photon coherent times can be produced by lasers. % \cite{Yariv}.
To summarize this brief comparison, we notice that the differences
between phonons and photons can often hinder our effort to apply ideas
originating in quantum optics to phonons.  For example, there does not
exist good phonon cavities at the moment, so that it is very difficult
to select phonon modes.  For phonons, we have to almost always deal
with a continuum of phonon modes.  In addition, short phonon lifetimes
and the strongly dissipative environment of phonons also complicate the
problems.  These will be taken into consideration when we work on the
theory.
\section{Phonon Operators and the Phonon Vacuum and Number States}
A phonon with quasimomentum ${\bf p} = \hbar {\bf q}$ and branch
subscript $\lambda$ has energy $\epsilon_{{\bf q}\lambda} = \hbar
\omega_{{\bf q} \lambda}$; the corresponding creation and annihilation
operators satisfy the boson commutation relations
\begin{equation}
\left[ b_{{\bf q'}\lambda'},\ b^{\dagger}_{{\bf q}\lambda} \right] =
\delta_{ {\bf q} {\bf{q'} } }\delta_{\lambda \lambda'} \,, \ \ \ \
\left[ b_{{\bf q}\lambda},\ b_{{\bf q'}\lambda'} \right] = 0 \,.
\end{equation}
The atomic displacements $u_{i\alpha}$ of a crystal lattice are given
by
\begin{equation}
u_{i\alpha} = \frac{1}{\sqrt{Nm}} \sum^{N}_{{\bf q} \lambda}
U^{\lambda}_{{\bf q} \alpha}Q^{\lambda}_{\bf q} e^{i {\bf q} \cdot {\bf
R}_i} \,.
\end{equation}
Here ${\bf R}_i$ refers to the equilibrium lattice positions, $\alpha$
to a particular direction, and $Q^{\lambda}_{\bf q}$ is the phonon
normal-mode operator
\begin{equation}
Q^{\lambda}_{\bf q} = \sqrt{\frac{\hbar}{2 \omega_{{\bf q} \lambda}} }
\left( b_{{\bf q} \lambda} + b^{\dagger}_{-{\bf q} \lambda} \right)
\,.
\end{equation}
For simplicity, hereafter we will drop the branch subscript $\lambda$,
assume that $U_{{\bf q} \alpha}$ is real, and define a ${\bf q}$-mode
dimensionless lattice amplitude operator:
\begin{equation}
u(\pm {\bf q}) = b_{\bf q} + b^{\dagger}_{-\bf q} + b_{-\bf q} +
b^{\dagger}_{\bf q} \,.
\end{equation}
This operator contains essential information on the lattice dynamics,
including quantum fluctuations.  It is the phonon analog of the
electric field in the photon case.
When no phonon is
excited, the crystal lattice is in the phonon vacuum state
$|0\rangle$.  The expectation values of the atomic displacement and the
lattice amplitude are zero, but the fluctuations will be finite:
\begin{eqnarray}
\langle(\Delta u_{i\alpha})^2 \rangle_{\rm vac} & \equiv & \langle
(u_{i\alpha})^2 \rangle_{\rm vac} - \langle u_{i\alpha} \rangle^2_{\rm
vac}  \\
& = & \sum_{\bf q}^{N} \frac{\hbar |U_{{\bf q}
\alpha}|^2}{2Nm\,\omega_{{\bf q}\alpha} } \,, \\
\langle(\Delta u(\pm {\bf q}))^2\rangle_{\rm vac} & = & 2 \,.
\end{eqnarray}
The eigenstates of the
harmonic phonon Hamiltonian are number states which satisfy $b_{\bf
q}|n_{\bf q}\rangle = \sqrt{n_{\bf q}} |n_{\bf q}-1\rangle$.  The
phonon number and the phase of atomic vibrations are conjugate
variables.  Thus, due to the uncertainty principle, the phase is
arbitrary when the phonon number is certain, as it is the case with any
number state $|n_{\bf q}\rangle$.  Thus, in a number state, the
expectation values of the atomic displacement $\langle n_{\bf
q}|u_{i\alpha}|n_{\bf q} \rangle$ and ${\bf q}$-mode lattice amplitude
$\langle n_{\bf q}|u(\pm {\bf q})|n_{\bf q} \rangle$ vanish due to the
randomness in the phase of the atomic displacements.  The fluctuations
in a number state $|n_{\bf q}\rangle$ are
\begin{eqnarray}
\langle(\Delta u_{i\alpha})^2\rangle_{\rm num} & = & \frac{\hbar
|U_{{\bf q} \alpha}|^2 n_{\bf q} }{Nm\,\omega_{{\bf q}\alpha} } +
\sum_{{\bf q'} \neq {\bf q}}^{N} \frac{\hbar |U_{{\bf q'}
\alpha}|^2}{2Nm\,\omega_{{\bf q'}\alpha} }  \,, \\
\langle(\Delta u(\pm {\bf q}))^2\rangle_{\rm num} & = & 2 + 2n_{\bf q}
\,.
\end{eqnarray}
\section{Phonon Coherent States}
A single-mode (${\bf q}$) phonon coherent state is an eigenstate of a
phonon annihilation operator:
\begin{equation}
b_{\bf q}|\beta_{\bf q}\rangle = \beta_{\bf q} |\beta_{\bf q} \rangle
\,.
\end{equation}
It can also be generated by applying a phonon displacement operator
$D_{\bf q}(\beta_{\bf q})$ to the phonon vacuum state
\begin{eqnarray}
|\beta_{\bf q} \rangle & = & D_{\bf q}(\beta_{\bf q}) |0\rangle =
\exp(\beta_{\bf q} b_{\bf q}^{\dagger} - \beta_{\bf q}^* b_{\bf
q})|0\rangle  \\
& = & \exp \left( -\frac{|\beta_{\bf q}|^2}{2} \right) \sum_{n_{\bf
q}=0}^{\infty} \frac{\beta_{\bf q}^{n_{\bf q}}}{\sqrt{n_{\bf q}!}}
|n_{\bf q}\rangle \,.
\end{eqnarray}
As is shown above, a phonon coherent state is a phase-coherent
superposition of number states.  Moreover, coherent states are a set of
minimum-uncertainty states which are as noiseless as the vacuum state.
Coherent states are also the set of quantum states that best describe
the classical harmonic oscillators \cite{Gardiner}.
A single-mode phonon coherent state can be generated by the Hamiltonian
\begin{equation}
H  =  \hbar \omega_{\bf q} \left( b^{\dagger}_{\bf q} b_{\bf q} +
\frac{1}{2} \right) + \lambda^*_{\bf q}(t) \, b_{\bf q} +
\lambda_{\bf q}(t) \, b_{\bf q}^{\dagger}
\label{eq:coherent-phonon}
\end{equation}
and an appropriate initial state.  Here $\lambda_{\bf q}(t)$ represents
the interaction strength between phonons and the external source.  More
specifically, if the initial state is a vacuum state, $|\psi (0)
\rangle = |0 \rangle$, then the state vector becomes a
single-mode coherent state thereafter
\begin{equation}
|\psi(t)\rangle  = |\Lambda_{\bf q}(t) \, e^{-i\omega_{\bf q}t}\rangle
\,,
\end{equation}
where
\begin{equation}
\Lambda_{\bf q}(t) =  - \frac{i}{\hbar} \int_{-\infty}^t \lambda_{\bf
q}(\tau) \, e^{i\omega_{\bf q}\tau}d\tau
\end{equation}
is the coherent amplitude of mode ${\bf q}$.  If the initial state is a
single-mode coherent state $|\psi(0)\rangle = | \alpha_{\bf q}
\rangle$, then the state vector at time $t$ takes the form
\begin{equation}
|\psi(t)\rangle = |\left\{ \Lambda_{\bf q}(t) + \alpha_{\bf q}
\right\} e^{-i\omega_{{\bf q}}t}\rangle \,,
\end{equation}
which is still a coherent state.
In a single-mode (${\bf q}$) coherent state $|\Lambda_{\bf q}(t) \,
e^{-i\omega_{\bf q}t} \rangle$, $\langle u_{i\alpha}(t) \rangle_{\rm
coh}$ and $\langle u(\pm {\bf q}) \rangle_{\rm coh}$ are sinusoidal
functions of time.  The fluctuation in the atomic displacements is
\begin{equation}
\langle(\Delta u_{i\alpha})^2\rangle_{\rm coh} = \sum_{\bf q}^{N}
\frac{\hbar |U_{{\bf q} \alpha}|^2}{2Nm\,\omega_{{\bf q}\alpha}} \,.
\end{equation}
The unexcited modes are in the vacuum state and thus all contribute to
the noise in the form of zero point fluctuations.  Furthermore,
\begin{equation}
\langle(\Delta u(\pm {\bf q}))^2\rangle_{\rm coh} = 2 \,.
\end{equation}
>From the expressions of the noise $\langle (\Delta u_{i \alpha })^2
\rangle_{\rm coh}$ and $\langle (\Delta u[\pm {\bf q})]^2 \rangle_{\rm
coh}$, it is impossible to know which state (if any) has been excited,
while this information is clearly present in the expression of the
expectation value of the lattice amplitude $\langle u(\pm {\bf q})
\rangle_{\rm coh}$.  These results can be straightforwardly generalized
to multi-mode coherent states.
Coherent phonons have been the subject of considerable interest in
recent years.  Typically, the dynamics of coherent
phonons are described by using {\it classical} equations of motion.
Here we present a {\it quantum} description and show that it is
consistent with the classical one and, as an additional bonus, contains
information on quantum fluctuations.
Coherent phonons can be generated by a femtosecond short pulse laser
\cite{xhurmp}.  A femtosecond pulse duration is much shorter than any
phonon period and therefore acts as a {\it delta-function driving
force}.  It can produce coherent longitudinal optical (LO) phonons
\cite{review,xhurmp}.  We can make a very simplified calculation by
replacing the coupling strength $\lambda_{\bf q}(t)$ with $A
\delta(t-t_0)$ in Eq.~(\ref{eq:coherent-phonon}), so that
\begin{equation}
H_{\rm coh} = \hbar \omega_{\bf q} b^{\dagger}_{\bf q} b_{\bf q} + A \,
\delta(t-t_0) \, b_{\bf q} + A^* \, \delta(t-t_0) \, b^{\dagger}_{\bf
q} \,.
\end{equation}
Here $A = |A|e^{i\phi_A}$ is a time-independent complex amplitude
containing the information of the photon-phonon interaction and the
coherent amplitude of the relevant modes in the incident optical
pulse.  We assume that the crystal is in the phonon vacuum state before
it is hit by the laser pulse at $t = t_0$.
The time-evolution operator $U (t, t_0)$ for the phonon mode can be
written as
\begin{eqnarray}
U (t, t_0) & = & \exp \left( -i \omega_{\bf q} t \; b^{\dagger}_{\bf q}
b_{\bf q} \right) \; \exp \left( - \frac{iA}{2 \hbar} b_{\bf q} e^{-i
\omega_{\bf q} t_0} - \frac{iA^*}{2\hbar} b_{\bf q}^{\dagger} e^{i
\omega_{\bf q} t_0} \right) \,.
\end{eqnarray}
In other words, for $t > t_0$, the crystal is in a coherent state
$|\Lambda_{\bf q} e^{-i\omega_{\bf q} t} \rangle$.  The coherent phonon
amplitude $\Lambda_{\bf q}$ is
\begin{equation}
\Lambda_{\bf q} = - \; \frac{i A^*}{2 \hbar} \, e^{i \omega_{\bf q}
t_0}.
\end{equation}
which is a constant complex number.  Thus, a very short laser pulse
conveniently provides a time-independent amplitude and a coherent phase
to the active phonon mode(s).  For $t > t_0$, the average of the
atomic displacement operator in the state $|\Lambda_{\bf q}
e^{-i\omega_{\bf q}t} \rangle \otimes |0_{-\bf q} \rangle$ becomes
\begin{eqnarray}
\langle u_{i\alpha} \rangle & = & \sqrt{\frac{\hbar}{2Nm \omega_{\bf
q}}} \left( U_{{\bf q} \alpha} \Lambda_{\bf q} e^{-i\omega_{\bf q}t}
e^{i {\bf q} \cdot {\bf R}_i} +  U_{{\bf -q} \alpha} \Lambda_{\bf q}^*
e^{i\omega_{\bf q}t} e^{-i {\bf q} \cdot {\bf R}_i} \right) \nonumber
\\
& = & -\sqrt{\frac{2\hbar}{Nm \omega_{\bf q}}} \left| U_{{\bf q}
\alpha} \Lambda_{\bf q} \right| \sin[(\omega_{q} (t - t_0) - {\bf q}
\cdot {\bf R}_i - \phi_A - \phi_U] \,.
\end{eqnarray}
These longitudinal optical phonons can have a coherence time of about
$50$ps at $10$K, and even longer at lower temperatures \cite{review}.
In the classical sense, ``coherent'' means a wave with a well-defined
phase, or waves that can interfere with each other when superimposed.
Here we have shown that a single-mode coherent state of phonons
generated by a short laser pulse is indeed a plane wave with a
well-defined phase.  Thus, these phonons in a {\it quantum} coherent
state are also coherent in a {\it classical} manner.  Furthermore, they
are in a minimum-uncertainty state.
\section{Phonon Squeezed States}
In order to reduce quantum noise to a level below the zero-point
fluctuation level, we need to consider phonon squeezed states.
Quadrature squeezed states are generalized coherent states
\cite{Walls1}.  Here ``quadrature'' refers to the dimensionless
coordinate and momentum.  Compared to coherent states, squeezed ones
can achieve smaller variances for one of the quadratures during certain
time intervals and are therefore helpful for decreasing quantum noise.
Figures~\ref{fig1} and~\ref{fig2} schematically illustrate several
types of phonon states, including vacuum, number, coherent, and
squeezed states.  These figures are the phonon analogs of the
illuminating schematic diagrams used for photons \cite{Walls1}.
A single-mode quadrature phonon squeezed state is generated from a
vacuum state as
\begin{equation}
|\alpha_{\bf q} , \xi \rangle = D_{\bf q}(\alpha_{\bf
q} ) S_{\bf q}(\xi ) |0\rangle \,;
\end{equation}
a two-mode quadrature phonon squeezed state is generated as
\begin{equation}
|\alpha _{{\bf q}_1}, \alpha _{{\bf q}_2}, \xi \rangle = D_{{\bf q}_1}
(\alpha _{{\bf q}_1})D_{{\bf q}_2} (\alpha _{{\bf q}_2}) S_{{\bf q}_1,
{\bf q}_2} (\xi) |0\rangle \,.
\end{equation}
Here $D_{\bf q}(\alpha_{\bf q})$ is the coherent state displacement
operator with $\alpha_{\bf q} = |\alpha_{\bf q}| e^{i\phi}$,
\begin{eqnarray}
S_{\bf q} (\xi ) & = & \exp \left( \frac{\xi^*}{2} b_{\bf q}^2 -
\frac{\xi}{2} b_{\bf q}^{\dagger \, 2} \right) \,, \\
S_{{\bf q}_1, {\bf q}_2} (\xi ) & = & \exp \left( \xi^* b_{{\bf q}_1}
b_{{\bf q}_2} - \xi b_{{\bf q}_1}^{\dagger} b_{{\bf q}_2}^{\dagger}
\right) \,,
\end{eqnarray}
are the single- and two-mode squeezing operator, and $\xi=r
e^{i\theta}$ is the complex squeezing factor with $r \geq 0$ and $ 0
\leq \theta < 2\pi$.  The squeezing operator $S_{{\bf q}_1, {\bf
q}_2}(\xi)$ can be produced by the following Hamiltonian:
\begin{eqnarray}
H_{{\bf q}_1, {\bf q}_2} & = & \hbar \omega_{{\bf q}_1}
b^{\dagger}_{{\bf q}_1} b_{{\bf q}_1} + \hbar \omega_{{\bf q}_2}
b_{{\bf q}_2}^{\dagger} b_{{\bf q}_2}  \nonumber \\
& & + \zeta(t) b^{\dagger}_{{\bf q}_1} b^{\dagger}_{{\bf q}_2} +
\zeta^*(t)  b_{{\bf q}_1} b_{{\bf q}_2} \,.
\end{eqnarray}
The time-evolution operator for such a Hamiltonian has the form
\begin{equation}
U(t) = \exp \left( - \frac{i}{\hbar} H_0 t \right) \exp \left[ \xi^*(t)
b_{{\bf q}_1}b_{{\bf q}_2} - \xi(t) b_{{\bf q}_1}^{\dagger} b_{{\bf
q}_2}^{\dagger} \right] \,,
\end{equation}
where
\begin{eqnarray}
H_0 & = &  \hbar \omega_{{\bf q}_1} b^{\dagger}_{{\bf q}_1} b_{{\bf
q}_1} + \hbar \omega_{{\bf q}_2} b^{\dagger}_{{\bf q}_2} b_{{\bf q}_2}
\,, \\
\xi(t) & = & \frac{i}{\hbar} \int_{-\infty}^t \zeta(\tau) \,
e^{i(\omega_{{\bf q}_1} + \omega_{{\bf q}_2}) \tau} d\tau \,.
\end{eqnarray}
Here $\xi(t)$ is the squeezing factor and $\zeta(t)$ is the strength of
the interaction between the phonon system and the external source; this
interaction allows the generation and absorption of two phonons at a
time.  The two-mode phonon quadrature operators have the form
\begin{eqnarray}
X({\bf q},{-\bf q}) & = & 2^{-3/2}\left( b_{\bf q} + b_{\bf
q}^{\dagger} + b_{-\bf q} + b_{-\bf q}^{\dagger} \right)   \\
& = & 2^{-3/2}u(\pm {\bf q}) \,, \nonumber \\
P({\bf q},{-\bf q}) & = & -i 2^{-3/2} \left( b_{\bf q} - b_{\bf
q}^{\dagger} + b_{-\bf q} - b_{-\bf q}^{\dagger} \right) \,.
\end{eqnarray}
We have considered two cases where squeezed states were involved in
modes $\pm {\bf q}$.  In the first case, the system is in a two-mode
($\pm {\bf q}$) squeezed state $|\alpha_{\bf q}, \alpha_{-\bf q}, \xi
\rangle$, ($\xi = re^{i\theta}$), and its fluctuation is
\begin{equation}
\langle [\Delta u(\pm {\bf q})]^2 \rangle = 2 \left( e^{-2r} \cos^2
\frac{\theta}{2} + e^{2r} \sin^2 \frac{\theta}{2} \right) \,.
\end{equation}
In the second case, the system is in a single-mode squeezed state
$|\alpha_{\bf q}, \xi \rangle$ ($\alpha_{\bf q} = |\alpha_{\bf
q}|e^{i\phi}$) in the first mode and an arbitrary coherent state
$|\beta_{-\bf q} \rangle$ in the second mode.  The fluctuation is now
\begin{eqnarray}
\langle [\Delta u(\pm {\bf q})]^2 \rangle & = & 1 + e^{2r} \sin^2
\left(\phi + \frac{\theta}{2} \right)  \nonumber \\
& & + e^{-2r}\cos^2 \left(\phi + \frac{\theta}{2} \right) \,.
\end{eqnarray}
In both of these cases,
$\langle [\Delta u(\pm {\bf q})]^2 \rangle$ can
be smaller than in coherent states (see Fig.~2).
\section{Second Order Raman Scattering (SORS)}
So far we have focused on concepts and the actual quantum mechanical
states relevant to coherent and squeezed phonons.  Now we will focus on
one particular approach to generate squeezed phonon states via second
order Raman scattering.
The SORS process originates from the quadratic term in
the polarizability change $\delta \! P_{\alpha \beta}$ of a crystal.
The photon-phonon interaction $V$ that leads to the SORS
process is \cite{Born}
$V = - \frac{1}{4} \sum_{\alpha \beta} \sum_{\bf q}^N \sum_{j j'}
P_{\alpha \beta}^{{\bf q}j, -{\bf q}j'} Q_{{\bf q}j} Q_{-{\bf q}j'}
E_{1\alpha} E_{2\beta} \,$.
Here, $E_{1\alpha}$ and $E_{2\beta}$ are electric field amplitudes
along $\alpha$ and $\beta$ directions with frequencies $\omega_1$ and
$\omega_2$.
The second-order polarizability tensor
$P_{\alpha \beta}^{{\bf q}j, -{\bf q}j'}$
satisfies
$P_{\alpha \beta}^{{\bf q}j, -{\bf q}j'} = P_{\alpha \beta}^{-{\bf q}j',
{\bf q}j} = P_{\alpha \beta}^{-{\bf q}j, {\bf q}j'} \,$.
Recall that the complex phonon normal mode operator $Q_{{\bf q}j}$ of
the phonons is related to the phonon creation $b_{-{\bf q}j}^{\dagger}$
and annihilation $b_{{\bf q}j}$ operators by
$Q_{{\bf q}j} = b_{{\bf q}j} + b_{-{\bf q}j}^{\dagger}$.
If the incident photon fields are not attenuated
we can treat the optical fields as classical waves, and
also consider the different pairs of $\pm {\bf q}$
modes as independent, and treat them separately.
Thus, for one particular pair of $\pm {\bf q}$ modes, the complete
Hamiltonian for the two phonon modes involved in the SORS
process has the form \cite{Born}:
$
{\cal H}_{\bf q}  =  H_{\bf q}
- \{ 4^{-1} \sum_{\alpha \beta}
P_{\alpha \beta}^{{\bf q}, -{\bf q}} E_{1\alpha} E_{2\beta} \}
Q_{\bf q} Q_{-\bf q} \; ,
$
\noindent where
$
H_{\bf q} = \hbar \omega_{\bf q}
\{ b_{\bf q}^{\dagger} b_{\bf q} +
       b_{-\bf q}^{\dagger} b_{-\bf q} \}
$
is the free phonon Hamiltonian for the modes ${\bf q}$ and $-{\bf q}$,
$\; \omega_{\bf q}= (\omega_1 - \omega_2 ) / 2 $, and the branch labels
$j$ and $j'$ have been dropped.
Here we consider two different cases.  The first is when the incident
photons are in two monochromatic beams \cite{xhurmp,xhuprl97}; i.e.,
with electric fields $E_j = {\cal E}_j \cos({\omega_j t}+\phi_j)$;
$j=1,\,2$.  In the second case the incident photons are in an
ultrashort pulse whose duration is much shorter than the phonon period
\cite{xhuprl97,Garrett}.
\section{Squeezed Phonons via Continuous Wave SORS}
Let us now first consider the continuous wave (CW) case.  Because the
photons are monochromatic, we can take a rotating wave approximation
\cite{Schubert} and keep only the on-resonance terms in the Hamiltonian.
The off-resonance terms only contribute to virtual processes
\cite{Srivastava} at higher orders.  This approximation is
appropriate for times much longer than the phonon period.
The simplified Hamiltonian has the form
\begin{eqnarray}
{\cal H}_{\bf q}^{(cw)} & = & H_{\bf q}
- \lambda_{\bf q} \left\{ b_{\bf q} b_{-\bf q} \,
e^{2i\omega_{\bf q} t + i\phi_{12}} + c.c.
\right\} \; , \nonumber \\
\lambda_{\bf q} & = & \frac{1}{16} | \sum_{\alpha \beta} P_{\alpha
\beta}^{{\bf q}, -{\bf q}} {\cal E}_{1\alpha} {\cal E}_{2\beta} | \,,
\end{eqnarray}
where $\phi_{12}$ and $\lambda_{\bf q}$ refer to the overall phase and
amplitude, respectively, of the product of the 2nd-order polarizability
and the incident electric fields.  Recall that $P_{\alpha \beta}^{{\bf
q}, -{\bf q}}$ is real, therefore the phase $\phi_{12}$ has no ${\bf
q}$-dependence.  It originates solely from the two photon modes.
The Schr\"{o}dinger equation for the $\pm {\bf q}$--mode phonons is
$
i \hbar \partial_t \left|\psi_{\bf q} (t)\right\rangle
= {\cal H}_{\bf q}^{(cw)} (t) \left|\psi_{\bf q} (t)\right\rangle \, ,
$
and its time-evolution operator can be solved by a transformation into
the interaction picture.  The result can be expressed as
\cite{xhurmp,xhuprl97}
\begin{equation}
\left|\psi_{\bf q} (t) \right\rangle  =
e^{ \{   H_{\bf q} \, t /  i\hbar  \} } \;
e^{ \{ \zeta_{\bf q}^* b_{\bf q} b_{-\bf q}
- \zeta_{\bf q} b_{\bf q}^{\dagger} b_{-\bf q}^{\dagger} \} }
\left|\psi_{\bf q} (0) \right\rangle \;, \\
\label{eq:cwstate}
\end{equation}
where $\zeta_{\bf q}  =  - \; i \lambda_{\bf q} \, t \, e^{-i\phi_{12}}
/ \hbar  \;$.  Notice that the second factor in the time-evolution
operator is a two-mode quadrature squeezing operator \cite{Loudon}.
In the CW case considered here, the amplitude of the squeezing factor
$\zeta_{\bf q}$ grows linearly with time.  However, this initial linear
growth will be eventually curbed by subsequent phonon-phonon scattering
and optical pump depletion.  In other words, the expression for the
squeezing factor $ \zeta_{\bf q} $ is valid for times much larger than
one phonon period, but much smaller than phonon lifetimes (because this
treatment considers non-decaying phonons).  By solving the phonon
Langevin equation, we have shown that the squeezing factor in the CW
SORS will eventually saturate at a constant value determined by the
strength of SORS and the phonon decay constant \cite{more}.  In
addition, the phase of the squeezing factor is determined by the phase
difference of the two incoming light waves.  If the $\pm {\bf q}$
phonon modes are initially in a vacuum state or in a coherent state,
the SORS will drive them into a two-mode quadrature squeezed state
\cite{xhurmp,xhuprl97}.
The time evolution operator of {\it all\/} the phonon mode pairs
(instead of just one pair of $\pm {\bf q}$ modes) that are involved
in this SORS process has the form
$
U(t) = \prod_{\bf q} U_{\bf q}(t) \,
$.
Therefore, as long as the photon depletion is negligible, all the
phonon modes that are involved in a SORS
process are driven into two-mode quadrature squeezed states.  In other
words, squeezing can be achieved in a continuum of phonon modes by a
CW stimulated SORS process.
\section{Squeezed Phonons via Impulsive SORS}
Recently, an impulsive SORS process has been used to experimentally
generate phonon squeezing\cite{Garrett}.  Here we treat the problem
expressing the time evolution operator of the system in terms of a
product of the two-mode quadrature squeezing operator and the free
rotation operators \cite{Schumaker}.  Since the incident photons are
now in an ultrashort pulse, the complete Hamiltonian
can be solved in the limit when the optical field can be
represented by a $\delta$-function.  Such an approximation is usually
considered when the optical pulse duration is much shorter than the
optical phonon period, which is experimentally feasible with
femtosecond laser pulses.  The Hamiltonian for the SORS can now be
written as
$
{\cal H}^{'}
=  \sum_{\bf q} \left\{
H_{\bf q}
- \lambda'_{\bf q} \delta(t) Q_{\bf q} Q_{-\bf q} \right\} \,$,
where $\lambda'_{\bf q}$ carries the information on the amplitudes of
the incoming optical fields and the electronic polarizability.  Notice
that the light-phonon coupling strength $\lambda_{\bf q}$ in the CW
case has units of energy, while $\lambda'_{\bf q}$ here has units of
$\hbar$.  To further simplify the problem, we assume that only $\pm
{\bf q}$ modes are involved in the process.  Such a simplification is
possible when the photon depletion and the phonon anharmonic
interaction are negligible, so that different pairs of phonon modes are
independent from each other.  The Hamiltonian is now
\begin{equation}
{\cal H}_{\bf q}^{'}
\, = \, H_{\bf q}
- \lambda'_{\bf q} \delta(t) Q_{\bf q} Q_{-\bf q} \,,
\label{eq:hpulse}
\end{equation}
and the Schr\"{o}dinger equation for these two phonon modes is
$
i \hbar \partial_t |\psi_{\bf q} (t)\rangle
\, = \,
{\cal H}_{\bf q}^{'} \, |\psi_{\bf q} (t)\rangle
$.
This equation can be solved by separating the free oscillator terms and
the two-phonon creation and annihilation terms. The resulting
time-dependent wavefunction is
\begin{eqnarray}
|\psi_{\bf q} (t)\rangle & = &
\exp{ \left\{ \frac{ t H_{\bf q} }{ i \hbar } \right\} }
\ \exp{ \left\{ \frac{ i \lambda'_{\bf q} H_{\bf q} }
{ \hbar^2 \omega_{\bf q} } \right\} }
\nonumber \\
& & \times \exp{ \left\{ \zeta_{\bf q}^{' \, *} b_{\bf q} b_{-\bf q} -
\zeta'_{\bf q} b_{\bf q}^{\dagger} b_{-\bf q}^{\dagger} \right\} }
|\psi_{\bf q} (0^-)\rangle \, .
\label{eq:implusestate}
\end{eqnarray}
Here $\zeta'_{\bf q} = -i\lambda'_{\bf q} \, e^{-i\lambda'_{\bf
q}/\hbar}/\hbar$.  Hence the effect of the optical pulse is clear:  it
first applies a two-mode quadrature squeezing operator on the initial
state, then rotates the state by changing its phase \cite{Schumaker}.
The state will then freely evolve after $t =0^+$.  This result is
consistent with Ref.~\cite{Garrett} where the time-evolution operator
is expressed in terms of real phonon normal mode operators \cite{Born},
instead of the complex ones used in this paper.  Notice that, in
contrast to the CW SORS, the phase of the squeezing factor $\zeta'$ for
the impulsive case is fixed by the intensity of the light pulse.
\section{Macroscopic Implications and Time-dependence of the Dielectric
Constant}
Now that we have obtained the phonon states for both the CW and pulsed
SORS cases, let us consider the macroscopic implications of these
states.  More generally, let us first discuss the implications of the
phonon squeezed states disregarding how they are generated.  An
experimentally observable quantity $O$ which is related to the atomic
displacements in the crystal can generally be expressed in terms of
$Q_{\bf q}$:
$ O = O(0) + \sum_{{\bf q}} (\partial O/\partial Q_{\bf q})
Q_{\bf q} + \ldots \, = O_0 + O_1 +  O_2 + \ldots $
where the first term $O_0 = O(0)$ is the operator $O$ when all $Q_{\bf
q}$'s vanish.  An example of an experimentally observable quantity $O$
is the change in the crystal dielectric constant $\delta \epsilon$ due
to the atomic displacement produced by the incident electric fields.
To first order in $Q_{\bf q}$,
$
\delta \epsilon = \delta \epsilon_1 = \sum_{q_x > 0} \left| \frac{\partial
( \delta \epsilon )}{\partial Q_{\bf q}} \right|
\sqrt{\frac{\hbar}{2\omega_{\bf q}}} ( b_{{\bf q}} + b_{{-\bf
q}}^{\dagger} ) e^{i\Psi_{\bf q}} + ( b_{{-\bf q}}
+ b_{{\bf q}}^{\dagger} ) e^{-i\Psi_{\bf q}} ] \,
$.
Here $\Psi_{\bf q}$ is the phase of $\partial O/\partial Q_{\bf q} =
\partial (\delta \epsilon)/\partial Q_{\bf q}$.  Indeed, a widely used
method to track the phases of coherent phonons in the time-domain
\cite{review} is based on the observation of the reflectivity (or
transmission) modulation $\delta R$ ($\delta T$) of the sample, which
is linearly related to $\delta \epsilon$---the change in the dielectric
constant due to lattice vibrations.
The above equation for $\delta \epsilon$ indicates that we can
define a generalized \cite{xhurmp} lattice amplitude operator
\cite{xhuprb,xhuprl}:
$ \
u_g (\pm {\bf q}) = \left( b_{\bf q} + b^{\dagger}_{-\bf q} \right)
e^{i\Psi_{\bf q}} + \left( b_{-\bf q} + b^{\dagger}_{\bf q} \right)
e^{-i\Psi_{\bf q}}  \,.
$
This generalized lattice amplitude $u_g (\pm {\bf q}) = 2 Re\{ Q_{\bf
q} e^{i\Psi_{\bf q}} \}$ is the underlying microscopic quantity related
to an observed reflectivity or transmission modulation when the linear
term in $Q_{\bf q}$, $\; \delta \epsilon_1$, exists.
Even if $\langle\delta\epsilon_1\rangle$ vanishes, $\langle \Delta
(\delta \epsilon_1)^2 \rangle$ does not.  Since different pairs of $\pm
{\bf q}$ phonon modes are uncorrelated to one another, the fluctuation
of $\delta \epsilon_1$ can be expressed as
$
\langle (\Delta \delta \epsilon_1)^2 \rangle = \sum_{q_x > 0} \left(
\hbar/2\omega_{\bf q} \right)
$
$
\left| \partial \left( \delta \epsilon
\right) / \partial Q_{\bf q} \right|^2 \langle \Delta u_g^2 (\pm {\bf
q}) \rangle \,
$.
Here the state is $|\psi(t)\rangle = U(t) |\psi(0)\rangle = \prod_{\bf
q} U_{\bf q} (t) |\psi_{\bf q} (0)\rangle$ in either the CW or the
impulsive case.  We can again focus on a single pair of $\pm {\bf q}$
modes.  In the CW case, using Eq.~(\ref{eq:cwstate}), the fluctuation is
\begin{eqnarray}
\langle \Delta u_g^2 (\pm {\bf q}) \rangle^{(cw)}  & = & 2 \{
e^{-2r_{\bf q}} \cos^2 ( \Omega_{\bf q} (t) + \phi_{12}/2)
\nonumber \\
& & + e^{ 2r_{\bf q}} \sin^2 ( \Omega_{\bf q} (t) + \phi_{12}/2 ) \} \, ,
\label{eq:cwfluc}
\end{eqnarray}
where
$r_{\bf q} = |\zeta_{\bf q}| = \lambda_{\bf q} \; t /\hbar \,$,
$\Omega_{\bf q} (t) = \omega_{\bf q} t + \pi / 4$,
and hereafter $\langle \ldots \rangle$ denotes an expectation value
on squeezed states, unless stated otherwise.
Therefore, at certain times, the fluctuation $\langle \Delta u_g^2 (\pm
{\bf q}) \rangle^{(cw)}$ can be smaller than $2$, which is the vacuum
fluctuation level.  Furthermore, all the pairs of phonon modes that are
driven by the stimulated SORS
process share the same frequency:
$\omega_{\bf q} = (\omega_1 - \omega_2)/2$.  Therefore, all
the fluctuations $\langle \Delta u_g^2 (\pm {\bf q}) \rangle^{(cw)}$
evolve with the same $\omega_{\bf q}$.  Notice that there is no
dependence on $\Psi_{\bf q}$ in the final expression of $\langle \Delta
u_g^2 (\pm {\bf q}) \rangle^{(cw)}$, and the squeezing factor phase
$\phi_{12}/2$ has no ${\bf q}$--dependence, all the pairs of modes
involved through the SORS share the same phase in their fluctuations.
Therefore there can be squeezing in the overall fluctuation $\langle
(\Delta \delta \epsilon_1)^2 \rangle^{(cw)}$.  Furthermore, the phase of
this overall fluctuation can be adjusted by tuning the phase difference
of the two incoming light beams.
In the impulsive case \cite{Garrett,review}, if the $\pm {\bf q}$-mode
phonons are driven into a squeezed vacuum state, the fluctuation in
$u_g (\pm {\bf q})$ is
\begin{eqnarray}
\langle \Delta u_g^2 (\pm {\bf q}) \rangle'
= 2 \{ e^{-2r'_{\bf q}} \cos^2 \Omega'_{\bf q} (t) +
e^{2r'_{\bf q}} \sin^2 \Omega'_{\bf q} (t) \} \,,
\label{eq:impulsefluc}
\end{eqnarray}
where $r'_{\bf q} = |\zeta'_{\bf q}| = \lambda'_{\bf q} /\hbar \, ,$
and $\Omega'_{\bf q} (t) = \Omega_{\bf q} (t) - r'_{\bf q} $.
Again, the squeezing will reveal itself through oscillations in
$\langle [\Delta (\delta \epsilon_1)]^2 ({\bf q}) \rangle'$ which is
proportional to $ \langle \Delta u_g^2 (\pm {\bf q}) \rangle'$.  Note
that these oscillations are essentially the same as the ones obtained
in the CW case.  However, now the squeezing factor is time-independent.
Also, the $t=0$
phase $\pi/4 - r'_{\bf q}$ in Eq.~(\ref{eq:impulsefluc})
is ${\bf q}$--dependent.
Eq.(\ref{eq:impulsefluc}) can be rewritten as
$
\langle \Delta u_g^2 (\pm {\bf q}) \rangle' = 2 \left\{ \cosh 2r'_{\bf q}
+ \sinh 2r'_{\bf q} \; \sin ( 2\omega_{\bf q} t - r'_{\bf q} ) \right\} \,.
$
For small $r'_{\bf q}$, this becomes
$\langle \Delta u_g^2 (\pm {\bf q}) \rangle' = 2 \{ 1 + 2 r^{' \,
2}_{\bf q} + 2 r'_{\bf q} \sin ( 2\omega_{\bf q} t - r'_{\bf q} ) \}$.
This expression has essentially the same form as the one
obtained in \cite{Garrett}:
$\langle Q_{\bf q}^2 (t) \rangle =
 \langle Q_{\bf q}^2 (0) \rangle
\{ 1 + 2 \xi_{\bf q}^2 + 2 \xi_{\bf q}
\sin( 2 \omega_{\bf q} t + \varphi_{\bf q}) \}$.
The small phase term $\varphi_{\bf q}$ is neglected
in \cite{Garrett} when computing transmission changes.
The difference in phases,
$r'_{\bf q}$ versus $\varphi_{\bf q}$,
% which is of second order in $r'_{\bf q}$,
is negligible in the limit of very small squeezing factor, and
originates from the different interaction Hamiltonians used here and in
\cite{Garrett}.  The interaction term in \cite{Garrett} is proportional
to $u_g^2(\pm {\bf q})$ with $\Psi_{\bf q} = 0$ (notice that their
$Q_{\bf q}$ is real and based on standing wave quantization
\cite{Born}).  Therefore, the interaction Hamiltonian in \cite{Garrett}
is (in our notation)
$
V \propto u_g^2(\pm {\bf q}) \propto 2 Q_{\bf q} Q_{-\bf q} + Q_{\bf
q}^2 + Q_{-\bf q}^2 \,
$.
However, the last two terms in this expression do not satisfy momentum
conservation, we thus did not include them and kept only $Q_{\bf q}
Q_{-\bf q}$ in our interaction term (this form is also used by
Ref.\cite{Born}).
When the linear perturbation $\delta \epsilon_1$ due to phonons
vanishes, such as in \cite{Garrett}, the second order correction
$O_2 (= \delta \epsilon_2)$
must be considered.  When the phonon states are modulated by a SORS, so
that the $\pm {\bf q}$ modes are the only ones which are correlated,
$
\delta \epsilon_2 = \sum_{{\bf q}} \frac{\partial^2 (\delta \epsilon) }
{\partial Q_{\bf q} \, \partial Q_{-\bf q}} Q_{\bf q} Q_{-\bf q} \,.
$
Let us first focus on one pair of $\pm {\bf q}$ modes in the CW case.
In a vacuum state,
$
\langle 0 | Q_{\bf q} Q_{-\bf q} | 0 \rangle = 1 \,
$;
while in a squeezed vacuum state $|0 \rangle_{\rm sq} $,
$
{ }_{\rm sq}\langle 0 | Q_{\bf q} Q_{-\bf q} | 0 \rangle_{\rm sq}
= \langle \Delta u_g^2 (\pm {\bf q}) \rangle^{(cw)}/2 \,
$,
with the right hand side given in Eq.~(\ref{eq:cwfluc}).
Therefore, the expectation value of $Q_{\bf q} Q_{-\bf q}$ in a
squeezed vacuum state is periodically smaller than its vacuum state
value.  Let us now include all the phonon modes that contribute to
$\delta \epsilon_2$.  In a vacuum state,
$
\langle 0| \delta \epsilon_2 |0 \rangle = \sum_{\bf q} \partial^2
\delta \epsilon/(\partial Q_{\bf q} \, \partial Q_{-\bf q}) \,
$.
On the other hand, in a squeezed vacuum state,
\begin{equation}
\langle \delta \epsilon_2 \rangle = \frac{1}{2}\sum_{{\bf q}}
\frac{\partial^2 (\delta \epsilon)}{\partial Q_{\bf q} \, \partial
Q_{-\bf q}} \langle \Delta u_g^2 (\pm {\bf q}) \rangle^{(cw)} \,.
\end{equation}
\noindent Since the phase $\phi_{12}/2$ has no ${\bf q}$--dependence,
contributions from the phonon modes sharing the same frequency add up
constructively.  It is thus possible that
$\langle \delta \epsilon_2 \rangle$ is periodically smaller than its
vacuum state value.
Similarly, in the impulsive case,
$
\langle \delta \epsilon_2 \rangle' = 2^{-1} \sum_{{\bf q}}
\frac{\partial^2 (\delta \epsilon)}{\partial Q_{\bf q} \, \partial
Q_{-\bf q}} \langle \Delta u_g^2 (\pm {\bf q}) \rangle' \, ;
$
However, the phase factor in $\langle \delta \epsilon_2 \rangle'$
has a ${\bf q}$--dependence through $r'_{\bf q}$, so that all the
phonon modes with the same $\omega_{\bf q}$ do {\it not\/} contribute
to $\langle \delta \epsilon_2 \rangle'$ synchronously.
In the CW SORS and in the very-small-$r'_{\bf q}$ limit
impulsive SORS the phase of the expectation value
$\langle Q_{\bf q} Q_{-\bf q}\rangle$ does not depend on ${\bf q}$;
this is crucial to the experimental observation of modulations
in the dielectric constant, because this ${\bf q}$--insensitivity
leads to constructive summations of all the ${\bf q}$ pairs involved.
Also, at a van Hove singularity a large number of modes contribute
to $\delta \epsilon_2$ with the same frequency and phase,
thus their effect is larger and easier to observe \cite{Garrett}.
\section{Squeezed Phonons via a finite-width SORS}
Real light pulses are not $\delta$-functions.
Therefore, we have also considered a SORS pumped by a
light pulse with a finite width
(smaller than the phonon period $T$)
instead of a $\delta$-function.
For a fixed peak height $I$, we find\cite{more}
that the optimal pulse width $T_{\rm p}^{\rm opt}$
that maximizes the squeezing effect satisfies
$T_{\rm p}^{\rm opt} \approx T / 4.4 $.
This calculation indicates that the experiments\cite{Garrett}
used a pulse width which is nearby the optimal value
($T / 4.4 \approx 300 / 4.4 $ fs $\approx 68$ fs $ \approx T_{\rm p}$).
The calculation\cite{more} can be summarized as follows.
First, in the impulsive Hamiltonian we replace the
$\delta$-function by a Gaussian with its
width $T_{\rm p}$ as a variational parameter.
Since now the Hamiltonian is time-dependent in the interaction picture,
we cannot directly integrate the Schr\"{o}dinger equation.  Instead,
we use the Magnus method
to obtain the time evolution operator and keep only the dominant
first term.  This approximation is valid when the pulse duration is
shorter than the phonon period.  We then calculate the width
$T_{\rm p}^{\rm opt} $
of the Gaussian that maximizes the squeezing factor.
For a constant peak intensity,
a pulse that is too narrow does not contain enough photons; while
it can be proven that a pulse which is too long
(i.e., with a width comparable to $T$),
attenuates the squeezing effect.
\section{Phonon Squeezing Mechanism}
% {\it Phonon Squeezing Mechanism.}---
What is the mechanism of phonon
squeezing in the SORS processes?  For the CW case,
the Hamiltonian is the same as an optical two-mode parametric
process\cite{Schubert}, with the low frequency interference of
the combined photon modes as the pump, the two phonon modes as
the signal and idler.  The frequencies of these modes satisfy
$\omega_{\bf q} + \omega_{-\bf q} =
\omega_{1} - \omega_{2}$.
The impulsive case is slightly different.  Although the
Hamiltonian is similar to a parametric process, the energy transfer
from the photons to the two phonon modes is instantaneous.  The
resulting phonon state is a two-mode quadrature squeezed vacuum state.
Indeed, a regular parametric process pumps energy into the signal and
idler modes gradually, while the impulsive SORS does it suddenly.
The correlation between the two phonon modes, and thus the squeezing
effect, is also introduced instantaneously.  Notice that this mechanism
is reminiscent of the frequency-jump mechanism proposed in
\cite{Janszky}.  In the impulsive SORS, the frequency of the phonon
modes has an ``infinite'' $\delta$-peak change at $t = 0$, while the

frequency-jump mechanism has finite frequency changes, and squeezing
there can be intensified by repeated frequency jumps at appropriate
times.  However, as it has been pointed out in \cite{Janszky}, a finite
frequency jump up immediately followed by an equal jump down results in
no squeezing at all.
\section{Experimental Search for Squeezed Phonons}
One group\cite{Garrett} has claimed to detect
squeezed phonons.  Regrettably, no other group
has succeeded in reproducing their results.
Recent calculations indicate that these
initial experiments\cite{Garrett} do not prove that
phonon squeezing has been achieved experimentally.
Here we discuss several important points to be
considered for the eventual future experimental
observation of a phonon squeezed state.
First of all, squeezing must refer to a phonon mode with a variance
that falls {\it below\/} the standard quantum limit.  This crucial
comparison has not been produced yet.  Thus, a reliable determination
of the vacuum noise level is necessary in order to establish if a
squeezing of the vacuum noise has indeed occurred.  Experiments in
quantum optics have an independent way to reliably obtain the noise
level of the vacuum: by using an independent light beam as a local
oscillator.  Currently, there is no such phase-sensitive detection
scheme for phonons.  However, as we point out later, a pump-probe
experiment may be able to establish such a criteria, but discretion
has to be applied.
Second, for phonons, with relatively low energy compared to photons, thermal
noise should always be considered in an experiment.  This is especially
important when the noise modulation factor is small, such as the case in
the reported experiment \cite{Garrett}, where the noise modulation
factor is only 0.0001\%.  To achieve total noise modulation, as can be
done through CW and impulsive SORS, is not equivalent to the squeezing
of the vacuum noise.  Only when the noise modulation factor is big
enough to overcome the thermal noise, then the quantum noise is suppressed.
Third, in that experiment, the squeezing factor is
obtained from the modulation of the light transmission
through a crystal.  However, this change in light
transmission describes the modulation of the {\it total \/}
fluctuations (both quantum and thermal) of the atomic
displacements.  Thus, by itself this modulation of the
noise is not a proof of quantum noise suppression, let
alone of squeezing {\it below\/} the vacuum noise level.
Fourth, the efficiency of the signal detection is a
crucial issue in photon squeezing experiments, but not
addressed in its phonon "analog" experiment.
In a homodyne or heterodyne detector for photons, if the efficiency of
the photon counters is less than one, additional noise is
introduced into the signal.  Similarly, in a pump-probe phonon detection
scheme, the probe light pulse and photodetector should introduce
additional noise into the final signal.  Therefore, a careful analysis
of these additional noise sources should be performed when employing a
pump-probe scheme to detect phonon squeezing.
For the above reasons, it is premature and unwarranted to
claim that squeezed phonons have been observed experimentally.
The evidence presented so far is incomplete and inconclusive.
The answer to the question:
``Can phonons be squeezed like photons?" is yes on theoretical
grounds, but the experimental proof still lies in the future.
The rest of this section presents quantitative
derivations in support of the statements made above.
Even with the reservations presented above regarding
the pump-probe scheme, we would like to point out that it does
provide a possible means for the detection of phonon squeezing.
Here we discuss the criteria for achieving squeezing of
quantum noise through an impulsive SORS process in a
pump-probe experiment.  We then give a numerical estimate
on whether squeezing of quantum noise has been achived in
\cite{Garrett}.
Experimentally, the observed quantity is the change in the transmission
$T$ due to the impulsive SORS process.  Up to second order in
$Q_{\bf q}$, $T$ can be expressed as
\[
T = T_0 + \sum_{\bf q} \frac{\partial T}{\partial Q_{\bf q}} \langle
Q_{\bf q} \rangle + \sum_{\bf q} \frac{\partial^2 T}{\partial Q_{\bf q}
\partial Q_{\bf q'}} \langle Q_{\bf q} Q_{\bf q'} \rangle \,,
\]
where the average is over the phonon states of the crystal.  In a
squeezed vacuum (or thermal) state considered here here,
$\langle Q_{\bf q} \rangle = 0$ and
$
\langle Q_{\bf q} Q_{-\bf q} \rangle = 2 \langle \Delta u^2 (\pm {\bf
q}, t) \rangle
$.
Therefore, $T$ does contain informations on the atomic
lattice fluctuations.
Through impulsive SORS, we can achieve modulation of the total noise
of the system, and the effect can be seen through a modulation of
the transmission $T$, or more specifically, $\Delta T / T$.
Starting from a {\it thermal} state, we have:
\begin{eqnarray*}
\frac{\Delta T}{T} & \approx & \frac{1}{T} \sum_{\bf q}
\frac{\partial^2 T}{\partial Q_{\bf q} \partial Q_{-\bf q}}
[1 + 2N(\omega_{\bf q})] r_{\bf q} \sin(2\omega_{\bf q}t) \,.
\end{eqnarray*}
Notice that this modulation of the {\it total\/} noise is not
equivalent to a squeezing of quantum noise, because $\Delta T = 0$ is
an indication of {\it total\/} noise limit (That is, quantum
noise plus thermal noise), {\it not quantum\/} noise limit.  Even
though the experiment is done at low temperatures, so that thermal
noise is miniscule, we still have to compare the thermal noise to the
squeezing factor (or modulation factor through impulsive SORS) because
the later is very small, too.  One criteria for {\it quantum} noise
suppression would be
\begin{eqnarray*}
\frac{\Delta T^{\prime}}{T} & = &
\frac{T - T_{\rm vac}}{T_{\rm vac}} \\
& \approx & \frac{1}{T_{\rm vac}} \sum_{\bf q} \frac{\partial^2
T}{\partial Q_{\bf q} \partial Q_{-\bf q}} \left[ 2N(\omega_{\bf q}) +
r_{\bf q} \sin(2\omega_{\bf q}t) \right] \,.
\end{eqnarray*}
A negative $\Delta T^{\prime}/T$ here indicates that, at least in some
spectral regions, the atomic displacement noise is squeezed below the
ground state limit.  We are now comparing the thermal population
$2N(\omega_{\bf q})$ and the squeezing factor $r_{\bf q}$.  In summary,
there is always modulation of total noise by the impulsive SORS, but
only when $r_{\bf q} > 2N(\omega_{\bf q})$ do we achieve noise smaller
than the ground state limit.
For example, at an experimental temperature of $10$K, the corresponding
thermal energy $k_B T$ is about $0.88$meV.  The acoustic phonons
involved in the experiment reported in \cite{Garrett} have a frequency
of about $2.7$THz, which corresponds to an energy quantum of $\hbar
\omega \approx 10.6$meV.  Therefore, the thermal noise factor here is
about $N = 1/\{\exp(\hbar \omega / k_B T) - 1 \} \approx e^{-12}
\approx 0.6*10^{-5}$, which is almost identical to the squeezing factor
given in \cite{Garrett}.  Therefore, it is quite clear that this
particular experiment has not achieved squeezing of {\it quantum} noise
below the vacuum noise limit.  However, if the experimental temperature
is further lowered, the thermal factor will become smaller.  The
squeezing action initiated by the second order Raman scattering should
then be strong enough to realize suppression of noise below the ground
state level.
\section{Conclusions}
We have presented an overview of the definitions relevant to
quantum phonon optics, including quantum coherent and squeezed
phonons.
Afterwards, we have studied theoretically the generation of phonon
squeezing using a stimulated SORS process.  In particular, we
calculated the time evolution operators of the phonons in two different
cases: when the incident photons are in monochromatic continuous waves,
and when they are in an ultrashort pulse.  The amplitude of the
squeezing factor initially increases with time and then saturates in
the CW SORS case, while it remains constant in the pulsed SORS case.
In addition, the $t = 0$ phase
of the squeezing factor in the CW SORS, $\phi_{12}$,
can be continuously adjusted by tuning the relative
phase of the two incoming monochromatic photon beams,
while for the pulsed SORS the phase
($\propto \lambda_{\bf q}^{\prime}$) of the squeezing factor
is determined by the amplitude of the incoming light pulse.
For both cases we calculated the quantum fluctuations of a generalized
lattice amplitude operator and the second order contribution to the
change in dielectric constant, which is measurable.  For the
finite-width impulsive case, we computed the optimal pulse width, in
terms of the phonon period, that maximizes the squeezing effect.
\section{Acknowledgements}
We acknowledge useful conversations with S.~Hebboul, S.~Tamura
and H.~Wang.  One of us (XH) acknowledges support from the US
Army Research Office.

\newpage
\begin{figure}[h]
\caption[]{Schematic diagram of the uncertainty areas
(shaded) in the generalized coordinate and momentum ($X({\bf q},{-\bf
q})$, $\,P({\bf q},{-\bf q})$) phase space of (a) the phonon vacuum
state, (b) a phonon number state, (c) a phonon coherent state, and (d)
a phonon squeezed state.  Here $X({\bf q},{-\bf q})$ and $P({\bf
q},{-\bf q})$ are the two-mode ($\pm {\bf q}$) coordinate and momentum
operators, which are the direct generalizations of their corresponding
single-mode operators.  Notice that the phonon coherent state has the
same uncertainty area as the vacuum state, and that both areas are
circular, while the squeezed state has an elliptical uncertainty area.
Therefore, in the direction parallel to the $\theta/2$ line, the
squeezed state has a smaller noise than both the vacuum and coherent
states.}
\label{fig1}
\end{figure}
\begin{figure}[h]
\caption[]{Schematic diagram of the time evolution of the expectation
value and the fluctuation of the lattice amplitude operator $u(\pm {\bf
q})$ in different states.  Dashed lines represent $\langle u(\pm {\bf
q}) \rangle$, while the solid lines represent the envelopes $\langle
u(\pm {\bf q}) \rangle \pm \sqrt{\langle [\Delta u(\pm {\bf q})]^2
\rangle}$.  (a) The phonon vacuum state $|0\rangle$, where $\langle
u(\pm {\bf q}) \rangle = 0$ and $\langle [\Delta u(\pm {\bf q})]^2
\rangle = 2$.  (b) A phonon number state $|n_{\bf q}, n_{-\bf
q}\rangle$, where $\langle u(\pm {\bf q}) \rangle = 0$ and $\langle
[\Delta u(\pm {\bf q})]^2 \rangle = 2(n_{\bf q} + n_{-\bf q}) + 2$.
(c) A single-mode phonon coherent state $|\alpha_{\bf q} \rangle$,
where $\langle u(\pm {\bf q}) \rangle = 2|\alpha_{\bf q}| \cos
{\omega_{\bf q}t}$ (i.e., $\alpha_{\bf q}$ is real), and $\langle
(\Delta u[\pm {\bf q})]^2 \rangle = 2$.  (d) A single-mode phonon
squeezed state $|\alpha_{\bf q}e^{-i\omega_{\bf q}t}, \, \xi(t)
\rangle$, with the squeezing factor $\xi(t) = re^{-2i\omega_{\bf q}t}$
and $r = 1$.  Here, $\langle u(\pm {\bf q}) \rangle = 2|\alpha_{\bf
q}| \cos {\omega_{\bf q}t}$, and $\langle [\Delta u(\pm {\bf q})]^2
\rangle = 2(e^{-2r} \cos^2 {\omega_{\bf q}t} + e^{2r}\sin^2{\omega_{\bf
q}t})$.  (e) A single-mode phonon squeezed state, as in (d); now the
expectation value of $u$ is $\langle u(\pm {\bf q}) \rangle =
2|\alpha_{\bf q}| \sin {\omega_{\bf q}t}$, (i.e.  $\alpha_{\bf q}$ is
purely imaginary), and the fluctuation $\langle [\Delta u(\pm {\bf
q})]^2 \rangle$ has the same time-dependence as in (d).  Notice that
the squeezing effect now appears at the times when the lattice
amplitude $\langle u(\pm {\bf q}) \rangle$ reaches its maxima, while in
(d) the squeezing effect is present at the times when $\langle u(\pm
{\bf q}) \rangle$ is close to zero.}
\label{fig2}
\end{figure}
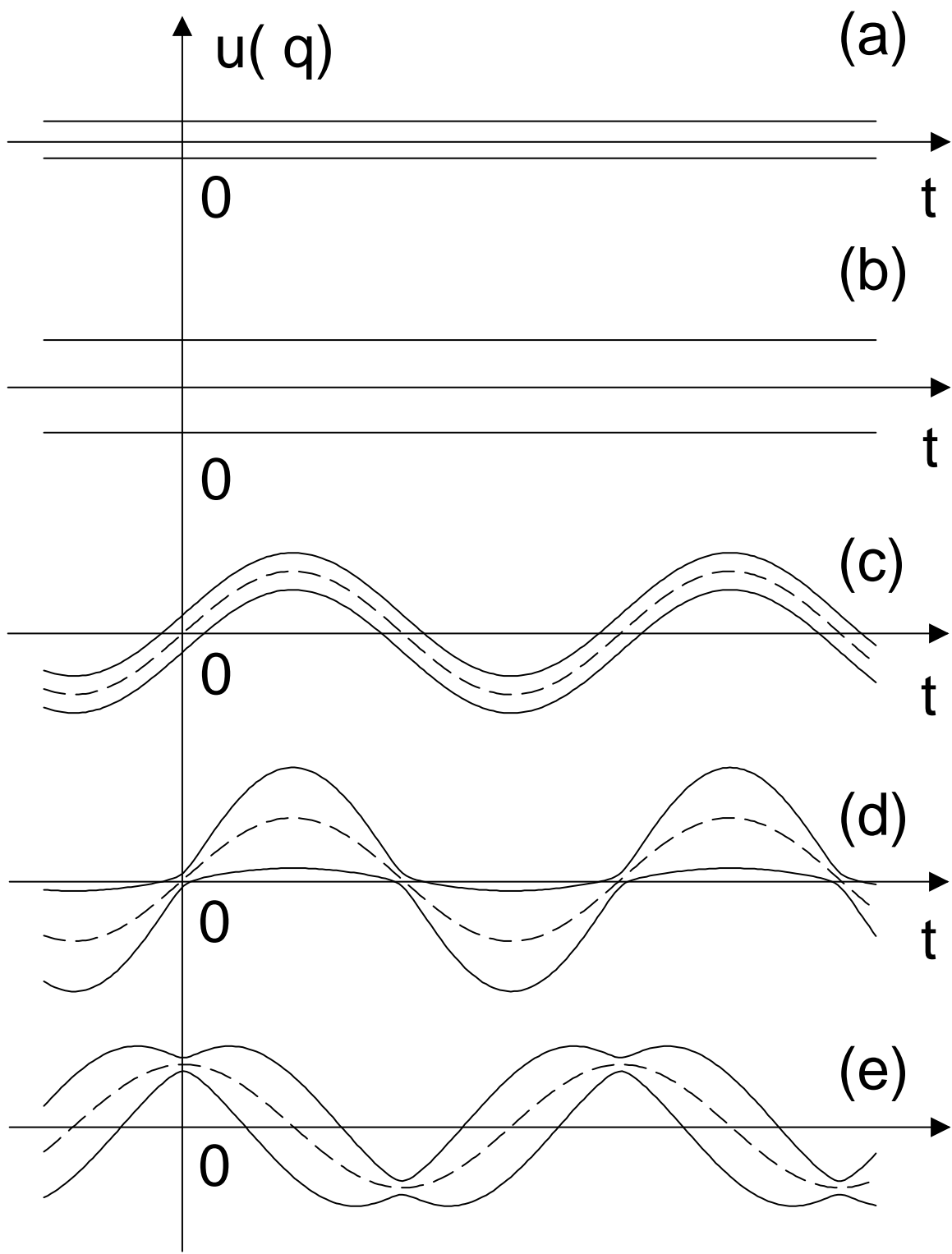
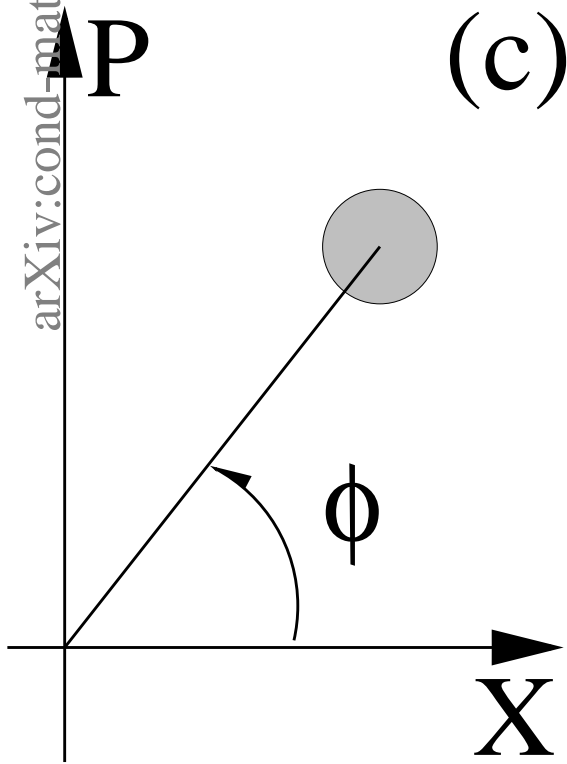
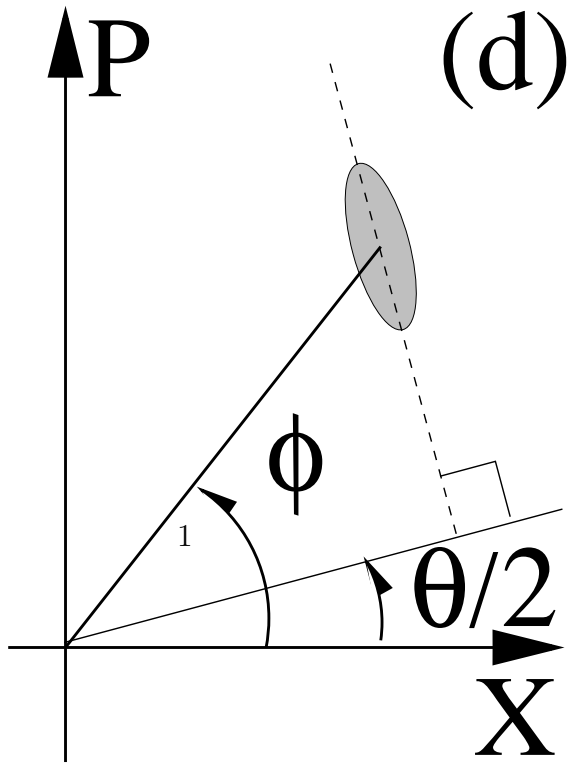
\end{document}